\documentclass[conference]{IEEEtran}
\IEEEoverridecommandlockouts
\usepackage{epsfig,rotating,setspace,latexsym,amsmath,epsf,amssymb,amsfonts,bm,theorem,cite,enumerate,longtable,accents,float,physics}
\usepackage{algorithm,algorithmic,graphicx,epsf,authblk,epstopdf,url,xcolor, soul,multirow,bbm}
\usepackage{mathtools,comment}
\usepackage[center]{qtree}
\usepackage{tree-dvips}
\usepackage[linguistics]{forest }

\newtheorem{theorem}{Theorem}
\newtheorem{corollary}{Corollary}
\newtheorem{definition}{Definition}
\newtheorem{remark}{Remark}
\newtheorem{lemma}{Lemma}
\allowdisplaybreaks

\newenvironment{Proof}[1]{\medskip\par\noindent{\bf Proof:\,}\,#1}{{\mbox{\,$\blacksquare$}\par}}

\usepackage{cite}
\usepackage{amsmath,amssymb,amsfonts}
\usepackage{algorithmic}
\usepackage{graphicx}
\usepackage{textcomp}
\usepackage{xcolor}
\newcommand{\RNum}[1]{\uppercase\expandafter{\romannumeral #1\relax}}

\def\BibTeX{{\rm B\kern-.05em{\sc i\kern-.025em b}\kern-.08em
    T\kern-.1667em\lower.7ex\hbox{E}\kern-.125emX}}
\begin{document}

\title{Convergence Properties of Good Quantum Codes for Classical Communication}

\author{Alptug Aytekin \qquad Mohamed Nomeir \qquad Lei Hu \qquad Sennur Ulukus\\
	\normalsize Department of Electrical and Computer Engineering\\
	\normalsize University of Maryland, College Park, MD 20742 \\
	\normalsize \emph{aaytekin@umd.edu} \qquad \emph{mnomeir@umd.edu} \qquad \emph{leihu@umd.edu} \qquad \emph{ulukus@umd.edu}}
\maketitle

\begin{abstract}
An important part of the information theory folklore had been about the output statistics of codes that achieve the capacity and how the empirical distributions compare to the output distributions induced by the optimal input in the channel capacity problem. Results for a variety of such empirical output distributions of good codes have been known in the literature, such as the comparison of the output distribution of the code to the optimal output distribution in vanishing and non-vanishing error probability cases. Motivated by these, we aim to achieve similar results for the quantum codes that are used for classical communication, that is the setting in which the classical messages are communicated through quantum codewords that pass through a noisy quantum channel. We first show the uniqueness of the optimal output distribution, to be able to talk more concretely about the optimal output distribution. Then, we extend the vanishing error probability results to the quantum case, by using techniques that are close in spirit to the classical case. We also extend non-vanishing error probability results to the quantum case on block codes, by using the second-order converses for such codes based on hypercontractivity results for the quantum generalized depolarizing semi-groups.
\end{abstract}

% \begin{IEEEkeywords} good codes, optimal output, strong converse.
% \end{IEEEkeywords}

\section{Introduction}
Han et al. \cite{tesunhan} were able to conclude an important conjecture regarding the output distribution of a good channel code, a code that has vanishing error probability while attaining near-capacity. It was priorly believed that for a good code for a discrete memoryless channel (DMC), the output distribution should resemble the optimal output distribution, which is attained by a sequence of i.i.d. random variables with the distribution induced by the optimal input distribution to the channel capacity maximization. They were able to show that, for a channel sequence $\textbf{W}$ that satisfies the strong converse, has a finite input alphabet and a finite capacity; a good code distribution should indeed satisfy
\begin{align}
    \lim_{n\to\infty}\frac{1}{n}D(\tilde{P}_{Y^n}\lVert P^*_{Y^n})=0, \label{hanverdu}
\end{align}
where $\tilde{P}_{Y^n}$ is the output distribution of the good code and $P^*_{Y^n}$ is the optimal output distribution. 

Shamai et al. \cite{shitz} then extended this result to show that the $k$th order empirical distribution of the good code, which is the type distribution for $k$-length strings, will satisfy a similar convergence result with respect to $k$-length optimal output sequence, as blocklength $n$ goes to infinity (and not the string length $k$). Moreover, they also derived convergence results for input distributions, by considering ``regular" good codes, codes whose input distributions are supported on the same support as an optimal input distribution. (Here it should be noted that the optimal output distribution is unique, as is shown in Lemma 1d of the same paper, however, as was shown in \cite{tesunhan}, the optimal input distribution is not unique.)

Polyanskiy et al. \cite{polyanskiy} then investigated another direction of this problem by considering whether such convergence results would hold for codes whose rates approach the capacity of the channel but have nonvanishing error for DMCs. Indeed, the proof for the convergence result of \eqref{hanverdu} follows by using Fano's inequality for the mutual information between the codeword distribution and the output distribution, and using the fact that the code has vanishing error probability to show that the mutual information approaches the capacity. Since the convergence would then be completely reliant on the convergence properties of the error (if there is any such property, as for general code sequences error does not need to be a function of the block length). However, Polyanskiy et. al were able to show that the convergence would still hold, with
\begin{align}
    D(\tilde{P}_{Y^n}\lVert P^*_{Y^n}) \leq nC-\log M_n+O(\sqrt{n}\log^{3/2}(n)) ,\label{anychan}
\end{align}
for any DMC, and for the special case of a DMC with no 0s on the stochastic matrix, this can be sharpened to 
\begin{align}
        D(\tilde{P}_{Y^n}\lVert P^*_{Y^n}) \leq nC-\log M_n+O(\sqrt{n}). \label{finitechan}
\end{align}
Moreover, they were able to show that for some ``nice" functions, the output distribution of the good code would cause a concentration around the expected value of the function under the optimal output distribution.

Here, it should be mentioned that Raginsky et al. \cite{ragisnky} were able to provide some sharpening on \eqref{anychan} and \eqref{finitechan} by explicitly showing the constants for the $O(\cdot)$ terms. It should also be mentioned that Liu et al. \cite{liu2019} were able to sharpen \eqref{anychan} even further, by showing that instead of $O(\sqrt{n}\log^{3/2}(n))$, it is still possible to get $O(\sqrt{n})$ for any DMC by using a technique which will be further explained below.

This kind of problem, and its rate-distortion version \cite{kanlis,weissman} have been well studied in the classical setting. In an effort to achieve similar ``empirical distribution" results for the transmission of classical information on a quantum channel, an analysis has been carried out in this paper by first showing that in our cases of interest, the optimal output distribution will exist and be unique, so that when we want to compare a good code's induced output distribution to the optimal one, there's no ambiguity regarding the optimal output distribution. Then, we are able to first extend Han et al. \cite{tesunhan} results for quantum codes for classical communication by using similar techniques to them, and then we are able to show \cite{polyanskiy}, or more precisely \cite{liu2019}, results for a subset of quantum codes by a modification of the techniques from \cite{cheng2019}.

\section{Problem Formulation}\label{problems}
Let $\mathcal{H}$ denote a Hilbert space, and $|\mathcal{H}|$ is the dimension of this Hilbert space. For our purposes, $|\mathcal{H}| < \infty$. When it is clear from context, $\mathcal{H}^n$ will be used to mean $\underbrace{\mathcal{H}\otimes \ldots \otimes \mathcal{H}}_{n\text{ times}}$.

$\mathcal{L}(\mathcal{H})$ is set of bounded linear operators of $\mathcal{H}$ and $\mathcal{D}(\mathcal{H})=\{\rho \in \mathcal{L}(\mathcal{H}):\ \rho=\rho^\dagger, ~ \rho \geq 0,~ Tr(\rho)=1\}$ is the set of density matrices, where the ordering is with respect to the positive semi-definite cone.

Positive operator-valued measure, POVM, denotes a set of operators $\{E_i\}$ such that $0 \leq E_i \leq I$ and $\sum_iE_i=I$. When a density matrix $\rho$ is measured under a POVM $\{E_i\}$, it means that outcome $i$ will happen with probability $Tr(\rho E_i)$, and afterwards, the density matrix will collapse. 

A quantum channel $\mathcal{N}:\mathcal{D}(\mathcal{H}_A)\to\mathcal{D}(\mathcal{H}_B)$ is a linear, completely-positive and trace-preserving map. Here, completely-positive means that for any arbitrary $\mathcal{H}_C$, and for any positive operator $V_{AC}\in \mathcal{H}_A\otimes\mathcal{H}_C$, $(\mathcal{N}_A\otimes I_C)(V_{AC})$ is a positive operator. When it is clear from context, $\mathcal{N}^n$ will be used to mean $\underbrace{\mathcal{N}\otimes \ldots \otimes \mathcal{N}}_{n\text{ times}}$.

A classical code for the quantum channel $\mathcal{N}$ is $(f,\{E_i\})$ where $\{E_i\}$ is POVM for decoding, $f:\{1,\ldots,M\}\to\mathcal{D}(\mathcal{H}_{A})$  is the encoder, and the overall state over all systems is given by
\begin{align}
    \omega=\frac{1}{M}\sum_{m=1}^{M}\sum_{a}p(a|m)\ket{m}\bra{m}\otimes\rho_{a}\otimes\mathcal{N}(\rho_a),
\end{align}
where $p(a|m)$ denotes the output distribution of encoder $f$ when the input is the $m$th message, and thus the summation is over all possible labels for the outputs of the encoder. If $p(a|m)=\delta_{a,a'}$ for some $a'$, the code is called deterministic.

$(f,\{E_i\})$ is called a $(n,M,\epsilon)$  classical code for the quantum channel $\mathcal{N}$ with
\begin{enumerate}
    \item maximal error probability $\epsilon$, if it holds true that 
    \begin{align}
        Tr((\mathcal{N}^n(f(m))E_m))\geq 1-\epsilon,\ \forall m\in\{1,\ldots,M\}
    \end{align}
    \item average error probability $\epsilon$, if it holds true that
    \begin{align}
        \frac{1}{M}\sum^{M}_{i=1}Tr\left(\mathcal{N}^n(f(m))E_m)\right)\geq 1-\epsilon.
    \end{align}
\end{enumerate}

\begin{remark}
    Notice that in our case, we did not put any restriction on $f(m)$, so that entangled states in $\mathcal{H}^n_A$ are allowed in $(n,M,\epsilon)$ classical code for quantum channel $\mathcal{N}$, which effectively means that there is no real distinction between $(1,M_n,\epsilon)$ codes for $\mathcal{N}^n$ and $(n,M_n,\epsilon)$ codes for $\mathcal{N}$.
\end{remark}

Define for any $0<\epsilon<1$,
\begin{align}
    \mathcal{C}_{\epsilon}(\mathcal{N})\triangleq\sup\{&R:\exists\delta \textrm{ s.t. } \forall n\geq n_0(\delta), \exists (n,M_n,\epsilon) \nonumber\\
    &\textrm{ code for $\mathcal{N}^n$ with } \frac{1}{n}\log M_n\geq R-\delta\}.
\end{align}

It was shown by Holevo, Schumacher and Westmoreland in \cite{Wilde} that $\mathcal{C}(\mathcal{N})=\underset{\epsilon\to 0^+}{\lim}{\mathcal{C}_{\epsilon}(\mathcal{N})}$ is given by
\begin{align}
    \mathcal{C}(\mathcal{N})&=\lim_{n\to\infty}\frac{1}{n}\chi(\mathcal{N}^n), \label{holevo} \\
    \chi(\mathcal{N}^n)&=\sup_{\substack{p_x,\rho_x: \\
    \sum_x p_x=1,\\
    \rho_x\in\mathcal{D}(\mathcal{H}_A)}}H\Big(\sum_{x}p_x\mathcal{N}^n(\rho_x)\Big)-\sum_{x}p_xH(\mathcal{N}^n(\rho_x)).
\end{align}

This theorem is the motivation for the following definition of ``good codes", that are codes that achieve the capacity.
\begin{definition}[Good codes]
    A sequence of classical codes for quantum channel $\mathcal{N}$ is called good codes if as $n\to\infty$, $\frac{1}{n}\log M_n\to C(\mathcal{N})$ and $\epsilon\to0$.
\end{definition} 

Channel $\mathcal{N}$ is said to satisfy the strong converse property, if $\mathcal{C}_\epsilon(\mathcal{N})=C, \ \forall0<\epsilon<1$. Moreover, note that $C=C(\mathcal{N})$. This special result motivates an extension of the definition of good codes.

\begin{definition}[Good codes for strong converse channels]
    A sequence of classical codes for quantum channel $\mathcal{N}$ satisfying strong converse property is called good codes if as $n\to\infty$, $\frac{1}{n}\log M_n\to C(\mathcal{N})$, without necessarily achieving vanishing error probability.
\end{definition}

\section{Preliminaries}\label{preliminaries}
\begin{definition}[Quantum relative entropy]
    Let $\rho,\sigma \in \mathcal{D}(\mathcal{H})$. The quantum relative entropy is defined as
    \begin{align}
        D(\rho\lVert\sigma)=\begin{cases}
            Tr(\rho\log\rho-\rho\log\sigma),\ \textrm{supp}(\rho)\subseteq\textrm{supp}(\sigma),\\
            +\infty,\ \textrm{otherwise},
        \end{cases}
    \end{align}
    where $\textrm{supp}$ denotes the support of a matrix, i.e. the complement of the kernel.
\end{definition}
The following lemma is useful as it can be used for showing convergence results by using entropic quantities. The following proof with ideas from \cite{Wilde, tomamichel} has been included here for completeness.
\begin{lemma}\label{same}
    For density matrices, $D(\rho\lVert\sigma)\geq0$, and moreover $D(\rho\lVert\sigma)=0$ iff $\rho=\sigma$. 
\end{lemma}
\begin{Proof}
    Using data processing inequality and choosing the trace out map as the operation, it can be seen that $D(\rho \lVert \sigma)\geq D(Tr(\rho)\lVert Tr(\sigma))=D(1\lVert 1)=0$.

    From the definition, it can be seen that when $\rho=\sigma$, $D(\rho \lVert \sigma)=0$. For the other way around, let $D(\rho \lVert \sigma)=0$, but $\rho \neq \sigma$. Data processing inequality then implies that $D(p_x||q_x)=0$ where $p_x=Tr(\rho E_x)$, $q_x=Tr(\sigma E_x)$, $\forall$ POVMs $\{E_x\}_x$. The strict convexity of $x\log x$ implies $D(\cdot\lVert \cdot)\geq 0$, and these combined implies that $D(p_x\lVert q_x)=0$ iff $p_x=q_x$ as otherwise we would have $D(p_x \lVert \frac{1}{2}p_x+\frac{1}{2}q_x) < 0$. This then says that $Tr(\rho E_x)=Tr(\sigma E_x)$ $\forall$ POVMs $\{E_x\}_x$. However, using the POVM $\{\{\rho \geq\sigma\},I-\{\rho \geq\sigma\}\}$ in which $\{\rho \geq\sigma\}$ is the projection to the positive eigenvalue space of the Hermitian operator $\rho-\sigma$, we see that $Tr(\{\rho \geq\sigma\}\left(\rho-\sigma\right))=0$ implies that $\rho-\sigma$ does not have any positive eigenvalues, and as $Tr(\rho-\sigma)=0$, it similarly does not have any negative eigenvalues either, so as all eigenvalues are $0$, it must mean that $\rho-\sigma=0$.  
\end{Proof}

The following lemma from \cite{optsig} is used extensively, whose proof is given for completeness.
\begin{lemma}\label{maxdistance}
    Let $\mathcal{N}:\mathcal{D}(\mathcal{H}_A)\to\mathcal{D}(\mathcal{H}_B)$, $|\mathcal{H}_A|<\infty$. Let $\{\bar{p}_x,\bar{\rho}_x\}$ be such that 
    \begin{align}
        \chi(\mathcal{N})=H\Big(\sum_x\bar{p}_x\mathcal{N}(\bar{\rho}_x)\Big)-\sum_x\bar{p}_xH(\mathcal{N}(\bar{\rho}_x)).
    \end{align}
    Define $\omega=\sum_{x}\bar{p}_x\mathcal{N}(\bar{\rho}_x)$. Then, it holds true that $D(\mathcal{N}(\rho)\lVert\omega) \leq \chi(\mathcal{N}),\ \forall\rho\in\mathcal{D}(\mathcal{H}_A)$ , and moreover $\bar{p}_x\neq0$ only if $D(\mathcal{N}(\bar{\rho}_x)\lVert\omega) = \chi(\mathcal{N})$.
\end{lemma}

\begin{Proof}
    For the existence of $\{\bar{p}_{x},\bar{\rho}_{x}\}$, one can refer to \cite{optsig} to see why it would indeed exist under the given assumptions.

    Note that by the definition of von Neumann entropy and quantum relative entropy,
    \begin{align}
        \chi(\mathcal{N})&=\max_{\substack{p_z,\rho_z:\\ \sum_zp_z=1, \\ \rho_z\in\mathcal{D}(\mathcal{H}_A)}} \chi(\mathcal{N},\{p_z,\rho_z\}), \\
        \chi(\mathcal{N},\{p_z,\rho_z\})&=H\Big(\sum_z{p}_z\mathcal{N}({\rho}_z)\Big)-\sum_z{p}_zH(\mathcal{N}({\rho}_z))\\
        &=\sum_zp_zD\Big(\mathcal{N}(\rho_z)\lVert\sum_zp_z\mathcal{N}(\rho_z)\Big). \label{restrictholevo}
    \end{align}
    
    Assume there exists $\rho_{x^*}$ such that $D(\mathcal{N}(\rho_{x^*})\lVert\omega) > \chi(\mathcal{N})$. Let $\omega'=\lambda\mathcal{N}(\rho_{x^*})+(1-\lambda)\sum_x\bar{p}_x\mathcal{N}(\bar{\rho}_x)=\lambda \mathcal{N}(\rho_{x^*})+(1-\lambda)\omega$. Define the difference in $\chi(\mathcal{N},\{p_z,\rho_z\})$ when $\{\lambda,\rho_{x^*}\}\cup\{(1-\lambda)\bar{p}_x,\bar{\rho}_x\}$ is used as opposed to $\{\bar{p}_x,\bar{\rho}_x\}$ being used as
    \begin{align}
        \Delta(\lambda)&\triangleq\lambda D(\mathcal{N}(\rho_{x^*})\lVert\omega')+(1-\lambda)\sum_x\bar{p}_xD(\mathcal{N}(\bar{\rho}_x)\lVert\omega')
        \nonumber\label{holevodifference}\\
        &\quad-\sum_x\bar{p}_xD(\mathcal{N}(\bar{\rho}_x)\lVert\omega) \\
        &=\lambda D(\mathcal{N}(\rho_{x^*})\lVert\omega')-\sum_x\bar{p}_xD(\mathcal{N}(\bar{\rho}_x)\lVert\omega) \nonumber\\
        &\quad+(1-\lambda)\bigg(\sum_x\bar{p}_xD(\mathcal{N}(\bar{\rho}_x)\lVert\omega)+D(\omega\lVert\omega')\bigg)\\
        &=\lambda \bigg(D(\mathcal{N}(\rho_{x^*})\lVert\omega')-\sum_x\bar{p}_xD(\mathcal{N}(\bar{\rho}_x)\lVert\omega)\bigg)\nonumber\\
        &\quad +(1-\lambda)D(\omega\lVert\omega')\\
        &\geq\lambda \bigg(D(\mathcal{N}(\rho_{x^*})\lVert\omega')-\sum_x\bar{p}_xD(\mathcal{N}(\bar{\rho}_x)\lVert\omega)\bigg) \\
        &=\lambda \left(D(\mathcal{N}(\rho_{x^*})\lVert\omega')-\chi(\mathcal{N})\right)\label{lowboundchange}.
    \end{align}

    Choose $\rho_n=\mathcal{N}(\rho_{x^*})$, $\omega'_{n}=\lambda(n) \mathcal{N}(\rho_{x^*})+(1-\lambda(n))\omega$ with $\lambda(n)\to 0$ as $n\to \infty$. Thus, $\rho_n\to\mathcal{N}(\rho_{x^*}),\ \omega'_n\to \omega$. Then, use the lower semicontinuity of the quantum relative entropy \cite{wehrl}, to get
    \begin{align}
        D(\mathcal{N}(\rho_{x^*})\lVert \omega) &\leq \liminf_n D(\rho_n \lVert\omega'_n)\\
        &=\sup_n\inf_{m\geq n}D(\rho_n \lVert \omega'_n)\\
        &=\sup_n\inf_{m\geq n}D(\mathcal{N}(\rho_{x^*}) \lVert \omega'_n).
    \end{align}
    The definition of supremum then implies that $\exists n$ such that $D(\mathcal{N}(\rho_{x^*}) \lVert \omega) \leq \inf\limits_{m\geq n}D(\mathcal{N}(\rho_{x^*}) \lVert \omega'_n)$. Using $\chi(\mathcal{N}) \leq D(\mathcal{N}(\rho_{x^*}) \lVert \omega)$, this in turn implies $\chi(\mathcal{N}) \leq D(\mathcal{N}(\rho_{x^*}) \lVert \omega'_m)$ for $\forall m\geq n$ through \eqref{holevodifference}. This shows the existence of $\lambda(m)>0$ such that $\Delta(\lambda(m))>0$. This in turn shows that $\{\{\lambda(m),\rho_{x^*}\},\{(1-\lambda(m))\bar{p}_x,\bar{\rho}_x\}\}$ is more optimal than 
    $\{\bar{p}_{x},\bar{\rho}_{x}\}$, a contradiction. Moreover, using \eqref{restrictholevo}, this also indicates that $\bar{p}_{x'}\neq0$ only if $D(\mathcal{N}(\bar{\rho}_{x'})\lVert\omega) = \chi(\mathcal{N})$.
\end{Proof}

\begin{lemma}\label{uniqueness}
    Let $\mathcal{N}:\mathcal{D}(\mathcal{H}_A)\to\mathcal{D}(\mathcal{H}_B)$, $|\mathcal{H}_A|<\infty$, and let $\{\bar{p}_{x},\bar{\rho}_{x}\}$ such that
    \begin{align}
       H\Big(\sum_{x}\bar{p}_{x}\mathcal{N}(\bar{\rho}_{x})\Big)-\sum_{x}\bar{p}_{x}H(\mathcal{N}(\bar{\rho}_{x})) =\chi(\mathcal{N}) < \infty,
    \end{align}
    and let $\{\tilde{p}_{z},\tilde{\rho}_{z}\}$ be another ensemble such that 
    \begin{align}
         H\Big(\sum_{z}\tilde{p}_{z}\mathcal{N}(\tilde{\rho}_{z})\Big)-\sum_{z}\tilde{p}_{z}H(\mathcal{N}(\tilde{\rho}_{z})) =\chi(\mathcal{N}).   
    \end{align}
    Then, $\omega=\sum_{x}\bar{p}(x)\mathcal{N}(\bar{\rho}_{x})=\sum_z\tilde{p}_z\mathcal{N}(\tilde{\rho}_z)$. 
    
    That is to say, for any ensemble $\{p_{x},\rho_{x}\}$ such that $H\left(\sum_{x}{p}_{x}\mathcal{N}({\rho}_{x})\right)-\sum_{x}{p}_{x}H(\mathcal{N}({\rho}_{x})) =\chi(\mathcal{N})$, $\omega=\sum_xp_x\mathcal{N}(\rho_x)$ is unique.
       
\end{lemma}
\begin{Proof}
    Let $\{\bar{p}_{x},\bar{\rho}_{x}\}$ and $\{\tilde{p}_{z},\tilde{\rho}_{z}\}$ both satisfy $\chi(\mathcal{N})$, and $\omega=\sum_x\bar{p}_x\mathcal{N}(\bar{\rho}_x), \ \tilde{\omega}=\sum_z\tilde{p}_z\mathcal{N}(\tilde{\rho}_z)$. From Lemma \ref{maxdistance}, note that for all $\{p_x,\rho_x\}$
    \begin{align}
        \chi(\mathcal{N}) \geq \sum_xp_xD(\mathcal{N}(\rho_x)\lVert\omega),
    \end{align}
    so that
    \begin{align}
        \chi(\mathcal{N}) &\geq \sum_z\tilde{p}_zD(\mathcal{N}(\tilde{\rho}_z)\lVert\omega) \\
        &=\sum_z\tilde{p}_zD(\mathcal{N}(\tilde{\rho}_z)\lVert\tilde{\omega})+D(\tilde{\omega}\lVert\omega) \\
        &=\chi(\mathcal{N})+D(\tilde{\omega}\lVert\omega),
    \end{align}
    which implies by finiteness of $\chi(\mathcal{N})$ that $D(\tilde{\omega}\lVert\omega)=0$, which from Lemma \ref{same} implies $\tilde{\omega}=\omega$.
\end{Proof}

\begin{remark}
    Lemma \ref{uniqueness} showing the uniqueness of optimal output state is indeed crucial in making sense in the following theorems; as without such a result stating that the optimal output state is unique, it would be much harder to talk about a convergence result of the output state induced by a code to optimal output state, as in that case such a state is not unique. In fact, assume the existence of two sequences of good codes, $\{\rho_m^n\}_m$ and $\{\sigma_m^n\}_m$, whose output distributions converge to two different optimal output states, say $\kappa_1$ and $\kappa_2$. Then, let 
    \begin{align}
        \!\omega_m^n=\begin{cases}
        \rho_m^n,\ \textrm{$n$ is odd},\\
        (\rho_{m_1,1}^{\frac{n}{2}},\sigma_{m_2,1}^{\frac{n}{2}},\rho_{m_1,2}^{\frac{n}{2}},\sigma_{m_2,2}^{\frac{n}{2}},\ldots),\ \textrm{$n$ is even},
        \end{cases}
    \end{align}
    where $\rho_{m}^n=(\rho^n_{m,1},\ldots,\rho^n_{m,n})$ and likewise for $\sigma_m^n$ and $m$ is split into two equal parts in even-numbered uses of the channel, i.e., $m=(m_1,m_2)$ in the usual manner of time-sharing. Time sharing in even-numbered uses will not decrease the rate by the fact that both $\{\rho_m^n\}_m$ and $\{\sigma_m^n\}_m$ are good codes. Thus, $\omega_m^n$ is a sequence of good codes. Notice that the output distribution of this sequence would not converge at all, as the output distribution of odd-numbered subsequences converges to $\kappa_1$, whereas the output distribution of even-numbered sequences converges to $\frac{\kappa_1}{2}+\frac{\kappa_2}{2}$.
\end{remark}

\section{Results}\label{results}
\subsection{Asymptotic Properties of Good Codes}
\begin{theorem}\label{qhanverdu}
    Let $\mathcal{N}:\mathcal{D}(\mathcal{H}_A)\to\mathcal{D}(\mathcal{H}_B)$ be a quantum channel such that $|\mathcal{H}_A|<\infty$. Let $(n,M_n,\epsilon)$ be a good code sequence for the channel $\mathcal{N}$. Let $\{\bar{p}_{x^n},\bar{\rho}_{x^n}\}$ such that
    \begin{align}
        \chi(\mathcal{N}^n)=H\Big(\sum_{x^n}\bar{p}_{x^n}\mathcal{N}^n(\bar{\rho}_{x^n})\Big)-\sum_{x^n}\bar{p}_{x^n}H(\mathcal{N}^n(\bar{\rho}_{x^n})).
    \end{align}
    Then, it holds true that
    \begin{align}
        \lim_{n\to\infty}\frac{1}{n}D(\omega_n\lVert \bar{\omega}_n)=0,
    \end{align}
    where $\bar{\omega}_n=\sum_{x^n}\bar{p}_{x^n}\mathcal{N}^n(\bar{\rho}_{x^n})$, and $\omega_n=\frac{1}{M_n}\sum_{m=1}^{M_n}\sum_{a}p(a|m)\mathcal{N}^n(\rho_a)$ which is the output state induced by the good code.
\end{theorem}
\begin{Proof}
    From Lemma \ref{maxdistance}, for any $\{p_{x^n},\rho_{x^n}\}\textrm{ and }\omega'_n=\sum_{x^n}p_{x^n}\mathcal{N}^n(\rho_{x^n})$, we have
    \begin{align}
        &\sum_{x^n}\bar{p}_{x^n}D(\mathcal{N}^n(\bar{\rho}_{x^n})\lVert\bar{\omega}_n)-\sum_{x^n}p_{x^n}D(\mathcal{N}^n(\rho_{x^n})\lVert \omega'_n) \nonumber\\
        &\quad= \chi(\mathcal{N}^n)-\sum_{x^n}p_{x^n}D(\mathcal{N}^n(\rho_{x^n})\lVert \omega'_n)\\
        &\quad\geq \sum_{x^n}p_{x^n}D(\mathcal{N}^n(\rho_{x^n})\lVert\bar{\omega}_n)-\sum_{x^n}p_{x^n}D(\mathcal{N}^n(\rho_{x^n})\lVert \omega'_n) \\
        &\quad=D(\omega'_n\lVert \bar{\omega}_n) \label{softconv1}.
    \end{align}
    Define the following states
    \begin{align}
        \sigma_n&=\frac{1}{M_n}\sum_{m=1}^{M_n}\sum_{a}p(a|m)\ket{m}_M\bra{m}\otimes\mathcal{N}^n(\rho_a)_B \\
        \upsilon_n&=\!\frac{1}{M_n}\!\sum_{m=1}^{M_n}\!\sum_{a,\hat{m}}p(a|m)Tr(E_{n,\hat{m}}\mathcal{N}^n(\rho_a))\!\ket{m,\hat{m}}\!\bra{m,\hat{m}}_{MB} \notag\\
        &=\sum_{m=1}^{M_n}\sum_{\hat{m}}p(m,\hat{m})\ket{m,\hat{m}}\bra{m,\hat{m}}_{MB},
    \end{align}
    where $\{E_{n,\hat{m}}\}_{\hat{m}}$ are the POVM decoding elements for $(n,M_n,\epsilon)$ code. Thus, note that $\upsilon_n=(I_M\otimes\mathcal{M}_B)(\sigma_n)$ where $\mathcal{M}$ is the measurement by $\{E_{n,\hat{m}}\}_{\hat{m}}$ channel. Also note that $p(m,\hat{m})$ is a probability distribution over $m,\hat{m}$. Further note that as $\upsilon_n$ is the state that is obtained by encoding and then decoding the messages according to the $(n,M_n,\epsilon)$ code that is being considered, thus it holds true that $p(\hat{m}=m)\geq 1-\epsilon$ (which is true whether the code follows maximal or average error criterion). Then,
    \begin{align}
        I(M;B)_{\upsilon_n}&=H(M)-H(M|B) \\
        &\geq \log(M_n)-(1+\epsilon\log M_n) \\
        &=(1-\epsilon)\log M_n -1
    \end{align}
    which follows from the usual Fano's inequality. Also note that
    \begin{align}
        I(M;B)_{\sigma_n}&=H(B)+H(M)-H(M,B) \\
        &=H\Big(\frac{1}{M_n}\sum_{m=1}^{M_n}\sum_ap(a|m)\mathcal{N}^n(\rho_a)\Big)+\log M_n\nonumber\\
        &\quad-\frac{1}{M_n}\sum_{m=1}^{M_n}H\Big(\sum_ap(a|m)\mathcal{N}^n(\rho_a)\Big)-\log M_n \\
        &= H\Big(\frac{1}{M_n}\sum_{m=1}^{M_n}\sum_ap(a|m)\mathcal{N}^n(\rho_a)\Big)\nonumber\\&\quad-\frac{1}{M_n}\sum_{m=1}^{M_n}H\Big(\sum_ap(a|m)\mathcal{N}^n(\rho_a)\Big) \\
        &=\frac{1}{M_n}\sum_{m=1}^{M_n}D\Big(\sum_ap(a|m)\mathcal{N}^n(\rho_a)\Big\lVert\omega_n\Big).
    \end{align}
Then, using the data processing inequality which states that $I(M;B)_{\sigma_n} \geq I(M;B)_{\upsilon_n}$, we obtain
\begin{align}
    \frac{1}{M_n}\sum_{m=1}^{M_n}D\Big(\sum_ap(a|m)\mathcal{N}^n(\rho_a)\Big\lVert\omega_n\Big) \geq (1-\epsilon)\log M_n-1 \label{softconv2}.
\end{align}
Selecting $\rho_{x^n}=\sum_ap(a|m)\rho_a,\ p_{x^n}=\frac{1}{M_n}$ which means $\omega'_n=\omega_n$ then gives in \eqref{softconv1} that
\begin{align}
    D(\omega_n\lVert \bar{\omega}_n) &\leq \sum_{x^n}\bar{p}_{x^n}D(\mathcal{N}^n(\bar{\rho}_{x^n})\lVert\bar{\omega}_n)\nonumber\\&\quad-\frac{1}{M_n}\sum_{m=1}^{M_n}D\Big(\sum_ap(a|m)\mathcal{N}^n(\rho_a)\Big\lVert\omega_n\Big) \\
    &\leq \sum_{x^n}\bar{p}_{x^n}D(\mathcal{N}^n(\bar{\rho}_{x^n})\lVert\bar{\omega}_n)+1-(1-\epsilon)\log M_n \\
    &=\chi(\mathcal{N}^n)+1-(1-\epsilon)\log M_n,
\end{align}
and we have
\begin{align}
    \frac{1}{n}D(\omega_n\lVert \bar{\omega}_n) \leq \frac{1}{n}\chi(\mathcal{N}^n)+\frac{1}{n}-(1-\epsilon)\frac{1}{n}\log M_n
\end{align}
From the definition of good codes, as $n\to\infty$, $\frac{1}{n}\log M_n\to C(\mathcal{N})=\lim_{n\to\infty}\frac{1}{n}\chi(\mathcal{N}^n)$ and $\epsilon\to0$, thus we see
\begin{align}
    \lim_{n\to\infty}\frac{1}{n}D(\omega_n\lVert \bar{\omega}_n)=0.
\end{align}
\end{Proof}
\begin{remark}
    Note that when the channel is additive, Lemma \ref{uniqueness} implies that the optimal output distribution for $n$ use of the channel would be a product state. When combined with Lemma \ref{qhanverdu}, any good code will asymptotically have a product output distribution even if entangled codewords have been used, so that the the channel in a way is asymptotically entanglement-breaking to its good codes.
\end{remark}

\subsection{Properties of Good Codes for Strong Converse Channels}
\begin{definition}[Block codes]
    $(n,M_n,\epsilon)$ is called a block code, if for any codeword $\rho_m$, $m\in\{1,\ldots,M_n\}$, it is possible to decompose it as $\rho_m=\otimes_{i=1}^n\rho_{m,i}$ where $\rho_m\in\mathcal{D}(\mathcal{H}_{A}^n),\rho_{m,i}\in\mathcal{D}(\mathcal{H}_{A}),\ \forall i\in\{1,\ldots,n\}$, while satisfying the error criterion.
\end{definition}
The following theorem, which is a modified version of the second-order converse in \cite[Theorem \RNum{3}]{cheng2019}, will make it possible to obtain the asymptotic properties that we want to achieve. The proof is similar to \cite{cheng2019} and can be found in the Appendix.
\begin{theorem}\label{secondorder}
    Let $\mathcal{N}:\mathcal{D}(\mathcal{H}_A)\to\mathcal{D}(\mathcal{H}_B)$ be a quantum channel with $|\mathcal{H}_B|<\infty$. For any deterministic $(n,M_n,\epsilon)$ block code under maximal error criterion, it holds true that
    \begin{align}
        \log(M_n)&\leq I(M;B^n)_\rho+2\sqrt{n(|\mathcal{H}_B|-1)\log\left(\frac{1}{1-\epsilon}\right)}\nonumber\\&\quad+\log\left(\frac{1}{1-\epsilon}\right),
    \end{align}
    where $\rho=\sum_{m=1}^{M_n}\frac{1}{M_n}\ket{m}\bra{m}_{M}\otimes \mathcal{N}(\rho_m)_{B^n}$.
\end{theorem}
From hereinafter, $B^n$ in the subscript on $\mathcal{N}(\rho_m)_{B^n}$ will be dropped unless it is not clear from the context.

\begin{lemma}\label{strongconvergence}
    Let $\mathcal{N}:\mathcal{D}(\mathcal{H}_A)\to\mathcal{D}(\mathcal{H}_B)$ be a quantum channel with $|\mathcal{H}_B|<\infty$. Let $\{\bar{p}_{x^n},\bar{\rho}_{x^n}\}$ such that
    \begin{align}
        \chi(\mathcal{N}^n)=H\left(\sum_{x^n}\bar{p}_{x^n}\mathcal{N}^n(\bar{\rho}_{x^n})\right)-\sum_{x^n}\bar{p}_{x^n}H(\mathcal{N}^n(\bar{\rho}_{x^n})).
    \end{align}
    For any deterministic $(n,M_n,\epsilon)$ block code under the maximal error criterion, it holds true that
    \begin{align}
        D(\mathcal{N}^n(\tilde{\rho})\lVert\mathcal{N}^n(\bar{\omega})) &\leq \chi(\mathcal{N}^n)-\log M_n +\log(1-\epsilon)\nonumber\\&\quad-2\sqrt{n(|\mathcal{H}_B|-1)\log\left(\frac{1}{1-\epsilon}\right)}
    \end{align}
    where $\bar{\omega}=\sum_{x^n}\bar{p}_{x^n}\bar{\rho}_{x^n}$, and $\tilde{\rho}=\frac{1}{M_n}\sum_{m=1}^{M_n}\rho_m$ with $\rho_m$ indicating the codeword of message $m$.
\end{lemma}
\begin{Proof}
    The logic we follow is in the same spirit as in the proof of Theorem \ref{qhanverdu}, that is to use the Donald's identity \cite{Donald_1987}, the quantum variant of what is classically called ``golden formula'', in conjunction with a converse result regarding the mutual information between the output and the message. Then, we are using Theorem \ref{secondorder} instead of the quantum version of the Fano inequality employed in the previous proof. 
    
    Define $\rho=\sum_{m=1}^{M_n}\frac{1}{M_n}\ket{m}\bra{m}_{M}\otimes \mathcal{N}(\rho_m)_{B^n}$ where $\rho_m$ are the codewords for the $(n,M_n,\epsilon)$ block code. Then,
    \begin{align}
        D(\mathcal{N}^n(\tilde{\rho})\lVert\mathcal{N}^n(\bar{\omega}))&=\frac{1}{M_n}\sum_{m=1}^{M_n}D(\mathcal{N}^n(\rho_m)\lVert\mathcal{N}^n(\bar{\omega}))\nonumber\\&\quad-\frac{1}{M_n}\sum_{m=1}^{M_n}D(\mathcal{N}^n(\rho_m)\lVert\mathcal{N}^n(\tilde{\rho})) \\
        &\leq \chi(\mathcal{N}^n)-I(M;B_n)_\rho \\
        &\leq \chi(\mathcal{N}^n)-\log M_n +\log(1-\epsilon)\nonumber\\&\quad-2\sqrt{n(|\mathcal{H}_B|-1)\log\left(\frac{1}{1-\epsilon}\right)}.\!
    \end{align}
\end{Proof}

The following corollary is a simple consequence of Lemma \ref{strongconvergence} and the Holevo-Schumacher-Westmoreland theorem, which shows the existence of $n$-block codes that achieve the single-letter capacity of the channel asymptotically, under the maximal error criterion \cite{Wilde}.

\begin{corollary}
    Let $\mathcal{N}:\mathcal{D}(\mathcal{H}_A)\to\mathcal{D}(\mathcal{H}_B)$ be a quantum channel with $|\mathcal{H}_B|<\infty$, $\chi(\mathcal{N}^n)=n\chi(\mathcal{N}), \forall n$, and satisfying the strong converse property. Let $\{\bar{p}_{x^n},\bar{\rho}_{x^n}\}$ such that
    \begin{align}
        \chi(\mathcal{N}^n)=H\Big(\sum_{x^n}\bar{p}_{x^n}\mathcal{N}^n(\bar{\rho}_{x^n})\Big)-\sum_{x^n}\bar{p}_{x^n}H(\mathcal{N}^n(\bar{\rho}_{x^n})).
    \end{align} 
    Then, for any good, deterministic $(n,M_n,\epsilon)$ block code under maximal error criterion, it holds true that
    \begin{align}
        \lim_{n\to\infty}\frac{1}{n}D(\mathcal{N}^n(\tilde{\rho})\lVert\mathcal{N}^n(\bar{\omega}))=0,
    \end{align}
    where $\bar{\omega}=\sum_{x^n}\bar{p}_{x^n}\bar{\rho}_{x^n}$, and $\tilde{\rho}=\frac{1}{M_n}\sum_{m=1}^{M_n}\rho_m$.
\end{corollary}

\begin{remark}
    To motivate the usefulness of the above corollary, we can give examples to the additive channels with strong converse property. Such channels include depolarizing, entanglement breaking, and Hadamard channels \cite{konig,Wilde,Wilde_2014,king}.
\end{remark}
\begin{remark}
    For general channels with strong converses, it is unknown whether there exists block codes that achieve the capacity of the channel under the nonvanishing error; whereas it is known that for vanishing error probability and for channels which are non-additive in Holevo information, block codes cannot reach the capacity.
\end{remark}
\begin{remark}
    As classical channels are a special case of quantum channels, using the counterexamples in \cite{polyanskiy}, we can indeed see that Lemma \ref{strongconvergence} cannot hold in general for stochastic encoders or average error criterion.
\end{remark}

\bibliographystyle{ieeetr}
\bibliography{reference.bib}

\section{Appendix}
\subsection{Proof of Theorem \ref{secondorder}}
The proof is in the same spirit as the one in \cite{cheng2019}, with instead of thinking about the classical-quantum channels, our focus is on the product states in order to focus on block codes.

The technique of the proof uses the weighted $L_p$ norms, and the contraction properties of such norms under operations. To that end, we define the weighted $L_p$ norm of $\rho$ with respect to $\sigma$ as
\begin{align}
    \lVert \rho \lVert_{p,\sigma}=Tr^{\frac{1}{p}}\left(\left|\sigma^{\frac{1}{2p}}\rho\sigma^{\frac{1}{2p}}\right|^{p}\right).
\end{align}

For $T\in\mathcal{D}(\mathcal{H}_B)$, define $\Phi_{t,\mathcal{N}(\rho_{m,i})}(T)=e^{-t}T+(1-e^{-t})Tr(\mathcal{N}(\rho_{m,i})T)I$, and define $\Psi(T)=e^{-t}T+(1-e^{-t})Tr(T)I$. Thus, in a manner similar, for $E_n\in\mathcal{D}(\mathcal{H}^n_B)$, define
\begin{align}
    \Phi_{t,\mathcal{N}^n(\rho_m)}(E_n)&=\left(\Phi_{t,\mathcal{N}(\rho_{m,1})}\otimes\ldots\otimes\Phi_{t,\mathcal{N}(\rho_{m,n})}\right)\left(E_n\right), \\
    \Psi_{t}(E_n)&=\left(\Psi_{t}\otimes\ldots\otimes\Psi_{t}\right)\left(E_n\right).
\end{align}

Define $D_\alpha^E(\rho \lVert \sigma)=\sup_{\{E_x\}_x}\frac{1}{\alpha-1}\log Q_{a}^E(\rho\lVert\sigma)=\frac{1}{\alpha-1}\log\left(\sum_xTr\left(E_x\rho\right)^{a}Tr\left(E_x\sigma\right)^{1-\alpha}\right)$ as the measured $\alpha$-Renyi relative entropy where the supremum is over all POVMs $\{E_x\}_x$. Let $\hat{\alpha}=\frac{\alpha}{\alpha-1}$ be the Hölder conjugate of $\alpha$. Using the equivalence of measured relative entropy to the projectively-measured relative entropy, and the variational expression found for the projectively-measured relative entropy by \cite{Berta2017} for $0<\alpha<1$, it holds true that
\begin{align*}
    D_{\alpha}^E(\rho \lVert\sigma)=\sup_{\omega>0}\frac{1}{\alpha-1}\log\left(Tr^{\alpha}(\rho\omega) Tr^{1-\alpha}(\sigma\omega^{\hat{\alpha}})\right).
\end{align*}

Note that $D^E_{1-\alpha}(\rho\lVert\sigma)=-\frac{\alpha-1}{\alpha}D_{\alpha}^E(\sigma\lVert\rho)$. Using this, with the variational expression by replacing $\rho=\mathcal{N}^n(\rho_m),\ \sigma=\frac{1}{M}\sum_{i=1}^m\mathcal{N}^n(\rho_m)=\mathcal{N}^n(\tilde{\rho})$, and $0<\alpha<\frac{1}{2}$, we have
\begin{align}
    &D_{1-\alpha}^E(\mathcal{N}^n(\rho_m)\lVert\mathcal{N}^n(\tilde{\rho})) \nonumber\\
    &\quad =-\frac{1}{\alpha}\inf_{\omega>0}\log\left(Tr^{1-\alpha}(\mathcal{N}^n(\rho_m)\omega^{{\hat{\alpha}}}) Tr^{\alpha}(\mathcal{N}^n(\tilde{\rho})\omega)\right) \\
    &\quad\geq-\frac{1}{\alpha}\log\Big(Tr^{1-\alpha}\left(\mathcal{N}^n(\rho_m)\Psi_t^n(E_{n,m})^{{\hat{\alpha}}}\right) \nonumber\\
    &\quad\quad \times Tr^{\alpha}\left(\mathcal{N}^n(\tilde{\rho})\Psi^n_t(E_{n,m})\right)\Big) \\
    &\quad=-\frac{1}{\alpha}\log\left(Tr^{1-\alpha}\left(\mathcal{N}^n(\rho_m)\Psi_t^n(E_{n,m})^{{\hat{\alpha}}}\right)\right)\nonumber\\
    &\quad\quad-\frac{1}{\alpha}\log\left(Tr^{\alpha}\left(\mathcal{N}^n(\tilde{\rho})\Psi^n_t(E_{n,m})\right)\right) \\
    &\quad=\log\left(Tr^{\frac{\alpha-1}{\alpha}}\left(\mathcal{N}^n(\rho_m)\Psi_t^n(E_{n,m})^{{\hat{\alpha}}}\right)\right)\nonumber\\
    &\quad\quad-\log\left(Tr\left(\mathcal{N}^n(\tilde{\rho})\Psi^n_t(E_{n,m})\right)\right), \label{termstobound}
\end{align}
where $\{E_{n,m}\}_m$ are the POVM elements for the $n$-block code (Here, the expression has in \cite{Berta2017} has been slightly modified, by changing $\omega\to\omega^{\frac{1}{\hat{\alpha}}}$, which is well-defined and an invertible operation as $\omega>0$).

Now, lower bounds for both terms in \eqref{termstobound} will be found. 

We start with the first term in \eqref{termstobound}. Note that $0<-\hat{\alpha}<1$, and $\frac{\alpha-1}{\alpha}<0$. Now, we use Araki-Lieb inequality for $-\hat{\alpha}$,
\begin{align}
    &Tr^{\frac{\alpha-1}{\alpha}}\left(\mathcal{N}^n(\rho_m)\Psi_t^n(E_{n,m})^{{\hat{\alpha}}}\right)\nonumber\\
    &\quad\geq Tr^{\frac{\alpha-1}{\alpha}}\left(\left(\mathcal{N}^n(\rho_m)^{-\frac{1}{2\hat{\alpha}}}\Psi_t^n(E_{n,m})^{-1}\mathcal{N}^n(\rho_m)^{-\frac{1}{2\hat{\alpha}}}\right)^{-{\hat{\alpha}}}\right) \\
    &\quad=\lVert \Psi^n_t(E_{n,m})^{-1}\lVert_{-{\hat{\alpha}},\mathcal{N}^n(\rho_m)}^{-1}
\end{align}
Now, we use the fact that $\Psi^n_t(E_{n,m})\geq \Phi_{t,\mathcal{N}^n(\rho_m)}(E_{n,m})$, to get $\Psi^n_t(E_{n,m})^{-1}\leq \Phi_{t,\mathcal{N}^n(\rho_m)}(E_{n,m})^{-1}$ and also the fact that $\lVert X \lVert_{p,\rho}\leq \lVert Y\lVert_{p,\rho}$ for $p>0$ and $\rho$, to get
\begin{align}
    &Tr^{\frac{\alpha-1}{\alpha}}\left(\mathcal{N}^n(\rho_m)\Psi_t^n(E_{n,m})^{{\hat{\alpha}}}\right)\nonumber\\
    &\quad\geq\lVert \Psi^n_t(E_{n,m})^{-1}\lVert_{-\hat{\alpha},\mathcal{N}^n(\rho_m)}^{-1} \\
    &\quad\geq \lVert \Phi_{t,\mathcal{N}^n(\rho_m)}(E_{n,m})^{-1}\lVert_{-\hat{\alpha},\mathcal{N}^n(\rho_m)}^{-1} \\
    &\quad=\lVert \Phi_{t,\mathcal{N}^n(\rho_m)}(E_{n,m})\lVert_{\hat{\alpha},\mathcal{N}^n(\rho_m)} \\
    &\quad\geq\lVert E_{n,m}\lVert_{q,\mathcal{N}^n(\rho_m)}
\end{align}
where the last step follows from the tensorized hypercontractivity inequality that has been established in \cite{Beigi2020} and holds for $t\geq \log\frac{\hat{\alpha}-1}{q-1}$, but for current purposes let $1>q=1+(\hat{\alpha}-1)e^{-t}>0$. Also, again by Araki Lieb, we get
\begin{align}
    \lVert E_{n,m}\lVert_{q,\mathcal{N}^n(\rho_m)}&=Tr^{\frac{1}{q}}\left(\left(\mathcal{N}^n(\rho_m)^{\frac{1}{2q}}E_{n,m}\mathcal{N}^n(\rho_m)^{\frac{1}{2q}}\right)^q\right) \\
    &\geq Tr^{\frac{1}{q}}\left(\mathcal{N}^n(\rho_m)E_{n,m}^q\right) \\
    &\geq Tr^{\frac{1}{q}}\left(\mathcal{N}^n(\rho_m)E_{n,m}\right) \\
    &\geq (1-\epsilon)^{\frac{1}{q}}, \label{firstpiece}
\end{align}
where the second last line follows from $0\leq E_{n,m}\leq I$ and the last line is by the error criterion of the code.

Now, we consider the second term in \eqref{termstobound}. Note that $\Psi_t^n(E_{n,m}) \leq \Psi_t^n(I)$ as $\sum_{m=1}^{M_n}E_{n,m}\leq I$ and $\Psi_t^n$ is positivity-preserving as can be seen from its definition. Then,
\begin{align}
    &\frac{1}{M_n}\sum_{m=1}^{M_n}Tr\left(\mathcal{N}^n(\tilde{\rho})\Psi_t^n(E_{n,m})\right) \nonumber\\
    &\quad\leq \frac{1}{M_n}Tr\left(\mathcal{N}^n(\tilde{\rho})\Psi_t^n(I)\right) \\
    &\quad=\frac{1}{M_n}Tr\left(\mathcal{N}^n(\tilde{\rho})\Psi_t^{\otimes n}(I)\right) \label{tensorizationidentity} \\
    &\quad=\frac{1}{M_n}Tr\left(\mathcal{N}^n(\tilde{\rho})\left(e^{-t}+|\mathcal{H}_B|(1-e^{-t})\right)^nI\right) \\
    &\quad\leq \frac{1}{M_n}e^{(|\mathcal{H}_B|-1)tn} Tr\left(\mathcal{N}^n(\tilde{\rho})\right) \label{convexityalpha}\\
    &\quad\leq \frac{1}{M_n}e^{(|\mathcal{H}_B|-1)tn}, \label{secondpiece}
\end{align}
where \eqref{tensorizationidentity} follows from the fact that identity on $\mathcal{H}_B^n$ is tensor product of the identity on $\mathcal{H}_B$, and \eqref{convexityalpha} follows from the convexity of $x^\alpha$ for $\alpha>1$ and then using the first-order convexity condition with $y=e^t$ and $x=1$.

Now, combining \eqref{firstpiece} and \eqref{secondpiece} with averaging over all codewords in \eqref{termstobound}, and then the using data processing inequality for the Petz-Renyi entropy,
\begin{align}
     &\frac{1}{q}\log(1-\epsilon)-nt(|\mathcal{H}_B|-1)+\log M_n \nonumber\\
     &\quad\leq  \frac{1}{M_n}\sum_{m=1}^{M_n}D_{1-\alpha}^E(\mathcal{N}^n(\rho_m)\lVert \mathcal{N}^n(\tilde{\rho})) \\
     &\quad\leq \frac{1}{M_n}\sum_{m=1}^{M_n}D_{1-\alpha}(\mathcal{N}^n(\rho_m)\lVert \mathcal{N}^n(\tilde{\rho})).
\end{align}
Now, letting $\alpha\searrow 0$, we also get $\hat{\alpha}\nearrow 0, q=1-e^{-t}$, to get
\begin{align}
    &\frac{1}{M_n}\sum_{m=1}^{M_n}D(\mathcal{N}^n(\rho_m)\lVert \mathcal{N}^n(\tilde{\rho})) \nonumber\\
    &\quad\geq \frac{1}{1-e^{-t}}\log(1-\epsilon)-nt(|\mathcal{H}_B|-1)+\log M_n \\
    &\quad\geq (1+\frac{1}{t})\log(1-\epsilon)-nt(|\mathcal{H}_B|-1)+\log M_n,
\end{align}
where the last line follows from $\frac{1}{1-e^{-t}}=\frac{e^t}{e^t-1}$ and $t+1<e^t$, and $t>0$. Then, optimizing over $t>0$, by differentiation we see that optimal $t$ is obtained at $t^*=\sqrt{-\frac{\log(1-\epsilon)}{n(|\mathcal{H}_B|-1)}}$, which then gives
\begin{align}
   &\log M_n+\log(1-\epsilon)-2\sqrt{-\log(1-\epsilon)n(|\mathcal{H}_B|-1)}\nonumber\\
   &\quad \leq \frac{1}{M_n}\sum_{m=1}^{M_n} D(\mathcal{N}^n(\rho_m)\lVert \mathcal{N}^n(\tilde{\rho})) \\
   &\quad=I(M;B^n)_\rho,
\end{align}
concluding the proof.
\end{document}